\documentclass[12pt,aps,prd,preprint,showpacs,nofootinbib,superscriptaddress]{revtex4-1}
\usepackage{epsf}
\usepackage{epsfig}
\usepackage{graphics,slashed}
\usepackage{graphicx}
\usepackage{amssymb}
\usepackage{amsmath}

\newcommand{\nc}{\newcommand}
\nc{\postscript}[2]
{\setlength{\epsfxsize}{#2\hsize}\centerline{\epsfbox{#1}}}
\nc{\non}{\nonumber}
\nc{\hc}{\hbox {h.c.}} \nc{\re}{\hbox {Re}} 
\nc{\mev}{\hbox {MeV}} \nc{\gev}{\;\hbox {GeV}} \nc{\tev}{\;\hbox {TeV}}
\def\lsim{\mathrel{\raise.3ex\hbox{$<$\kern-.75em\lower1ex\hbox{$\sim$}}}}
\def\gsim{\mathrel{\raise.3ex\hbox{$>$\kern-.75em\lower1ex\hbox{$\sim$}}}}

\nc{\etal}{{\it et al.}}
\nc{\Lsp}{\;\;\;\;\;\;\;\;\;\;}  \nc{\LLLsp}{\lspace \lspace}
\nc{\lsp}{\;\;\;\;\;\;}
\nc{\spac}{\;\;\;}
\nc{\noi}{\noindent}
\nc{\beq}{\begin{equation}}   \nc{\eeq}{\end{equation}}
\nc{\bea}{\begin{eqnarray}}   \nc{\eea}{\end{eqnarray}}
\nc{\baa}{\begin{array}}      \nc{\eaa}{\end{array}}
\nc{\bit}{\begin{itemize}}    \nc{\eit}{\end{itemize}}
\nc{\ben}{\begin{enumerate}}  \nc{\een}{\end{enumerate}}
\nc{\bce}{\begin{center}}     \nc{\ece}{\end{center}}

\def\n{{\nu}}

\def\sq2{\sqrt{2}}

\def\ph{\varphi}

\def\m4{m^4(\ph)}
\def\mn2{m_n^2}

\def\v5{V^{(5)}}

\begin{document}

\title{\begin{flushright}
       \mbox{\normalsize \rm DAWSON-HEP XX, CUMQ/HEP XXX}
       \end{flushright}
       \vskip 15pt
Phase broken $\mu-\tau$ symmetry and the neutrino mass hierarchy
}

\author{N.~Chamoun\footnote{Email: nidal.chamoun@hiast.edu.sy}}
\affiliation{ Physics Department, HIAST, P.O. Box 31983, Damascus, Syria.}
\author{C.~Hamzaoui\footnote{Email: hamzaoui.cherif@uqam.ca}}
\affiliation{
GPTP, D\'epartement des Sciences de la Terre et de L'Atmosph\`ere,
Universit\'e du Qu\'ebec \`a Montr\'eal, Case Postale 8888, Succ. Centre-Ville,
 Montr\'eal, Qu\'ebec, Canada, H3C 3P8}
\author{E.I.~Lashin\footnote{Email: slashin@zewailcity.edu.eg}}
\affiliation{Ain Shams University, Faculty of Science, Cairo 11566, Egypt.}
\author{S.~Nasri\footnote{Email: snasri@uaeu.ac.ae}}
\affiliation{Department of Physics, UAE University, P.O.Box 17551, Al-Ain, United Arab Emirates.}
\affiliation{International Center for Theoretical Physics (ICTP), Trieste, Italy.}
\author{M.~Toharia\footnote{Email: mtoharia@dawsoncollege.qc.ca}}
\affiliation{ Physics Department, Dawson College,
 3040 Sherbrooke St., Westmount, Quebec, Canada H3Z 1A4}

\date{\today}

\begin{abstract}

Inspired by the neutrino oscillations data, we consider the exact
$\mu-\tau$ symmetry, implemented at the level of the neutrino
mass matrix, as a good initial framework around which to study and describe neutrino
phenomenology. Working in the diagonal basis for the charged leptons, we deviate
from $\mu-\tau$ symmetry by just modifying the phases of the neutrino mass
matrix elements. This deviation is enough to allow for
a non-vanishing neutrino mixing entry $|V_{e3}|$ (i.e. $\theta_{13}$)
but it also gives a very stringent (and eventually falsifiable)
prediction for the atmospheric neutrino mixing element
$|V_{\mu3}|$ as a function of $|V_{e3}|$.
The breaking by phases is characterized by  a single phase and is
shown to lead to interesting lower bounds on the allowed mass of the lightest
neutrino depending on the ordering of neutrino masses (normal or
inverted) and on the value of the Dirac ${\cal CP}$ violating
phase $\delta_{CP}$. The allowed parameter space for the effective Majorana
neutrino mass $m_{ee}$ is also shown to be non-trivially constrained.

\end{abstract}

\maketitle

\section{Introduction}
\label{sec:intro}

Neutrinos are some of the most elusive particles of the Standard Model
(SM) since they interact mainly through weak processes. Nevertheless,
and thanks to the many succesful neutrino and collider experiments over the past
decades, we now have a pretty good understanding of the main features
of the lepton sector in particle physics.
Indeed we now know that neutrinos are massive but extremely light, and
that their individual masses are very similar. The leptonic mixing angles,
contrary to the quark mixing angles, are large and, in fact, the
relatively recent results from T2K \cite{T2K}, Double Chooz
\cite{DChooz},
RENO \cite{RENO} and Daya Bay \cite{DBay} Collaborations confirm that
even the angle $\theta_{13}$ of the neutrino mixing matrix is not that small.

We start this study with the observation that the data from
neutrino oscillations seem to show an approximate symmetry between the
second and third lepton families, also referred to as $\mu-\tau$
symmetry \cite{MUTAU,Harrison, MUTAU2} (see also \cite{Altarelli}).
Exact $\mu-\tau$ symmetry when implemented at the level of the Majorana
neutrino mass matrix $M_{\nu}$, leads to the following relations
between its elements, namely $M_{e \mu}=M_{e \tau}$ and $M_{\mu \mu}=M_{\tau \tau}$.
The neutrino mass matrix $M_{\nu}$ can thus be written as
\begin{eqnarray}
M_{\nu}^{\mu-\tau}=\begin{pmatrix}
m_{11} & m_{12} & m_{12} \cr
m_{12} & m_{22} & m_{23} \cr
m_{12} & m_{23} & m_{22} \end{pmatrix}\label{mutaumass}
\end{eqnarray}
where all entries are complex.
This particular texture, as well as different types of corrections
to it have been studied largely in the literature
\cite{literature}.
In particular, when implemented in the basis where the charged
lepton mass matrix $M_l$ is diagonal, the texture leads to the
vanishing of the mixing angle $|V_{e3}|$, which also implies a
vanishing of the Dirac measure of ${\cal CP}$ violation (even though
the phase $\delta_{CP}$ appearing in the usual PMNS parametrization remains
undefined), as well as a maximal atmospheric mixing element $|V_{\mu3}|=\frac{1}{\sqrt{2}}$.
The mixing matrix can then be described with a single free parameter
(to be fixed experimentally by the solar neutrino mixing element).
\begin{eqnarray}
V_{\mu-\tau} = \begin{pmatrix}{\scriptstyle \cos(\theta)} & {\scriptstyle \sin(\theta)} &
0 \cr -\frac{\sin(\theta)}{\sqrt{2}} & \frac{\cos(\theta)}{\sqrt{2}} & -\frac{1}{\sqrt{2}}
\cr -\frac{\sin(\theta)}{\sqrt{2}} & \frac{\cos(\theta)}{\sqrt{2}} & \frac{1}{\sqrt{2}}
\end{pmatrix}\  P ,
\end{eqnarray}
where $P$ is a diagonal matrix containing the Majorana phases.
Note that we use a particular phase convention different from the PDG one.
Also note that the neutrino masses $|m_1|$, $|m_2|$ and $|m_3|$ remain
as free parameters in the limit of exact $\mu$-$\tau$ symmetry.
Nevertheless experimental data constrain the differences among these
masses squared, with two possible orderings, {\it Normal Hierarchy
  (NH)}  and {\it Inverted Hierarchy (IH)} such that we have only one
free neutrino mass parameter, i.e the
lightest mass eigenvalue:
\bit
\item Normal Hierarchy (NH) ($|m_1|$ lightest):
  $|m_2|=\sqrt{\Delta^2_{sol}+|m_1|^2}$\   and\ $|m_3|=\sqrt{\Delta^2_{atm} + |m_1|^2}$
\item Inverted Hierarchy (IH)  ($|m_3|$ lightest):
  $|m_2|=\sqrt{\Delta^2_{sol}+|m_1|^2}$\
  and\ $|m_1|=\sqrt{\Delta^2_{atm} + |m_3|^2}$
\eit

The latest neutrino mixing global best fits \cite{Valle1,Conchita} lead to
\begin{eqnarray}
  \label{exp1}
|V^{exp}_{e3}|^2_{\rm (NH)} & =& 0.0216^{+0.0008}_{-0.0007} \\
|V^{exp}_{e3}|^2_{\rm (IH)} & =& 0.0222^{+0.0007}_{-0.0008} \\
|V^{exp}_{e2}|^2\ \  & =& 0.313^{+0.020}_{-0.016}  \\
|V^{exp}_{\mu 3}|^2_{\rm (NH)} & =&0.535^{+0.020}_{-0.029}\\
|V^{exp}_{\mu 3}|^2_{\rm (IH)} & =&0.539^{+0.018}_{-0.030}\\
\delta_{CP}/\pi  &=& 1.21^{+0.21}_{-0.15}\ \ \ \  ({\rm NH}) \\
\delta_{CP}/\pi & =& 1.56^{+0.13}_{-0.15} \ \ \ \  ({\rm IH})
\end{eqnarray}
and
\begin{eqnarray}
  \label{exp2}
  \Delta^2_{sol}&=& \left(7.55^{+0.20}_{-0.16}\right)\times 10^{-5} {\rm eV^2}\\
  \Delta^2_{atm}&=& \left(2.50\pm 0.03  \right) \times 10^{-3} {\rm eV^2} \ \ {\rm for\ NH}\\
  \Delta^2_{atm}&=& \left(2.42^{+0.03}_{-0.04} \right) \times 10^{-3} {\rm eV^2} \ \ \ \ {\rm for\ IH}
\end{eqnarray}
We see that the predicted values by $\mu$-$\tau$ symmetry for $|V_{e3}|$ and
$|V_{\mu3}|$ are quite close to the experimental values. Nevertheless $|V_{e3}|$ is
clearly measured to be non-zero (albeit relatively small) and thus the symmetry
should be somehow modified or broken in a controlled way.

We propose to modify the symmetric structure of the neutrino mass
matrix of Eq.~(\ref{mutaumass}) by adding phases that will break the
exact $2-3$ permutation symmetry.

\section{Analysis of Phase breaking of $\mu-\tau$ Symmetry}
\label{sec:phasebreaking}

Within the paradigm of $\mu-\tau$ symmetry, we can implement minimal deviations
at the level of the effective neutrino mass matrix $M_{\nu}$ given in
Eq.(\ref{mutaumass}) by adding phases to its elements \cite{Mohapatra}
(while maintaining its complex
symmetric nature).\footnote{See also \cite{Ramond} for a similar
  approach of phase breaking but at the level of the mixing matrix in
  the context of tri-bimaximal mixing \cite{tribimax}.}
We will thus assume that the neutrino mass matrix takes the form
\begin{eqnarray}
  \label{phasebrokenmutau0}
  M_{\nu}\equiv
  \left(\begin{array}{ccc} M_{ee} & M_{e \mu} &  M_{e \mu} e^{i \alpha} \\
  M_{e \mu} & M_{\mu \mu} & M_{\mu \tau}\ \ \ \\
  M_{e \mu}e^{i\alpha} & M_{\mu \tau} & M_{\mu \mu}e^{i \beta}
  \end{array}\right)
\eea
where all $M_{ij}$ entries are complex.
However, it  will prove very useful to parametrize this same mass
matrix as
\begin{eqnarray}
  \label{phasebrokenmutau}
  M_{\nu}\equiv
  P_\sigma\ \left(\begin{array}{ccc} m_{11} & m_{12}e^{i\theta} &  m_{12}e^{-i\theta} \\
  m_{12}e^{i\theta} & m_{22}e^{i\theta} & m_{23} \\
  m_{12}e^{-i\theta} & m_{23} & m_{22}e^{-i\theta} \end{array}\right)\ P_\sigma,
\end{eqnarray}
where all $m_{ij}$ parameters are complex and such that
$|m_{ij}| = |M_{ij}|$. The matrix $P_\sigma = diag(1, e^{-i\sigma},
e^{i\sigma})$ is a diagonal phase matrix such that under this
parametrization we have traded the original phases $\alpha$ and $\beta$ for the phases
$\theta$ and $\sigma$ which are now the source of {\it $\mu-\tau$ permutation
  symmetry} breaking. The conversion from the matrix in Eq.~(\ref{phasebrokenmutau0}) to the matrix
in Eq.~(\ref{phasebrokenmutau}) is given by $\theta = (\beta - 2 \alpha)/2$ and $\sigma = (\beta -\alpha)/2$.
It is important to note that the phase $\sigma$ will not enter into any physical
observable, given that it only appears within the phase matrix
$P_\sigma$, so that only the phase $\theta$ will have observable consequences.
Note that we assume that the charged lepton mass matrix is diagonal,
and the phases coming from it can be absorbed into the phase matrix
$P_\sigma$, without adding any physical consequence to the setup (see
for example \cite{Grimus2012}).     

We emphasize here  that the breaking of $\mu-\tau$ symmetry by phases is not 
a perturbative deviation from exact $\mu-\tau$ symmetry, unless the
phase $\theta$ is small, which we do not assume here.
Therefore our setup is not a special case of perturbed $\mu-\tau$
textures (see for example \cite{referee, Chamoun}) in which the breaking parameters are considered small.

Although it is not straightforward to find a symmetry leading directly to the
phase broken $\mu$-$\tau$ pattern,
it is still possible to find examples of field theoretical
realizations leading to this class of patterns.
As a simple example, let's take a type II seesaw scenario
in which the matter fields responsible for neutrino mass
generation are charged under an assumed $S\times Z_2$ symmetry
responsible for a $\mu-\tau$ symmetric flavor structure. 
When the neutral components of the Higgs triplets  $H^0_a,a=1,2,3$
within the type II seesaw scenario acquire vevs $v_a$, the $\mu-\tau$ flavour
symmetry is broken generating the following form for $M_\n$ \cite{Chamoun} 
\bea
M_\n & =&
\left(
\begin {array}{ccc}
v_1 G^1_{11} &v_2 G^2_{12}+v_3 G^3_{13}& v_2 G^2_{12} - v_3 G^3_{13} \\
v_2 G^2_{12}+v_3 G^3_{13}& v_1 G^1_{22}& v_1 G^1_{23}\\
v_2 G^2_{12} - v_3 G^3_{13} & v_1 G^1_{23}& v_1 G^1_{22}
\end {array}
\right).
\label{C1patternform}
\eea
 where $G_{ij}^a$ are Yukawa coupling constants and the indices $i,j$
 are flavor indices. For example, assuming that the vevs $v_2$ and $v_3$ are real,
 we can see directly that with the simple flavor constraint
 $G^3_{13}= i G^2_{12}$ on these Yukawa couplings, we obtain
 $|M_{e\mu}|=|M_{e\tau}|$ and thus one can reproduce naturally the 
 phase broken $\mu-\tau$ form of Eq.~(\ref{phasebrokenmutau0}) (in
 general, the necesary constraint to reproduce {\it  phase broken 
 $\mu-\tau$} would be that $Arg(v_3 G^3_{13}) - Arg(v_2 G^2_{12}) = \pi /2$ within  $\pi$).

The phenomenological effects of this texture should depart
smoothly from the usual $\mu-\tau$ symmetry predictions, which
correspond to the limit $\theta \to 0$. We
have dubbed this ansatz as {\it phase-broken $\mu-\tau$ symmetry}
and it can also obviously be described by the two conditions $|M_{e  \mu}| =|M_{e \tau}|$
and $|M_{\mu \mu}|=|M_{\tau \tau}|$ on the elements of the neutrino mass matrix.
The benefits of the particular parametrization of Eq.(\ref{phasebrokenmutau}) is that
it naturally includes the case of {\it $\mu-\tau$ reflection symmetry}
\cite{Harrison, Grimus} (see also for example \cite{Valle2, Xing, ReflectionOther} for
more recent references and references therein) as a special example of {\it phase-broken
$\mu-\tau$ symmetry}.\footnote{We will refer to the usual $\mu-\tau$ symmetry as
  {\it $\mu-\tau$ permutation symmetry} (leading to the constraints $M_{e \mu}=M_{e \tau}$
and $M_{\mu \mu}=M_{\tau \tau}$) and on the other hand we will refer to {\it $\mu-\tau$
    reflection symmetry} to the symmetry that leads to $M_{e \mu}=M^*_{e \tau}$ and
  $M_{\mu \mu}=M^*_{\tau \tau}$, with the further requirement that $M_{\mu \tau}$ and
  $M_{ee}$ be real.}

In our study, we will first consider the general consequences of {\it
 phase broken $\mu-\tau$ symmetry}, and then we will focus on specific
examples within selected regions of parameter space.

As a first example,  we will analyze the well studied ansatz of
$\mu-\tau$ {\it reflection symmetry}, understood as a limiting case scenario of {\it phase broken} $\mu-\tau$
symmetry. Using this approach we will show new analytical results for this
popular scenario of $\mu-\tau$ {\it reflection symmetry}.

We then will relax the conditions of $\mu-\tau$ {\it reflection}, but
we will keep the value of $\delta_{CP}$ to be $-\pi/2$, and then allow the Majorana
phases to have any value consistent with the phase-broken $\mu-\tau$ constraints.

Finally we will consider the cases where the Dirac phase $\delta_{CP}$ takes values
within the preferred ranges emerging from global fits, within both Normal and Inverted hierarchies.


\subsection{{\it Phase-broken } $\mu$-$\tau$: the general case}
\label{sec:phenogeneral}

The symmetric neutrino mass matrix must
be diagonalized in order to go to the neutrino physical basis. The
relationship between the mass matrix and the mass eigenvalues is
\begin{eqnarray}
M_{\nu} =U_\nu D_\nu U^T_\nu \label{diagonalization}
\end{eqnarray}
where the unitary  matrix $U_\nu$ is given by
\bea
U_\nu=P_L V_{PMNS}  \label{Umatrix}
\eea
and where $D_\nu=diag (|m_1|,|m_2|,|m_3|)$ is a diagonal matrix with
positive definite elements, and where  $P_L=diag(e^{i \gamma_1},e^{i \gamma_2},e^{i
  \gamma_3})$ is an unphysical diagonal phase matrix.

Choosing $|V_{e2}|$, $|V_{e3}|$, $|V_{\mu 3}|$, $\delta_{CP}$, $\eta$
and $\xi$ as our 6 independent parameters, we shall parametrize the
$V_{PMNS}$ mixing matrix as,
\begin{eqnarray}
V_{PMNS}= \left(\begin{array}{ccc} |V_{e1}| & |V_{e2}|e^{i\eta} & |V_{e3}|e^{i(\xi-\delta_{CP})} \\
V_{\mu1} & V_{\mu2} & -|V_{\mu3}|e^{i\xi} \\ V_{\tau1} & V_{\tau2} &
|V_{\tau3}|e^{i\xi} \end{array}\right)
\label{vpmns}
\end{eqnarray}
where $\delta_{CP}$ is the so-called Dirac phase and $\eta$ and $\xi$ are the
so-called Majorana phases and where the rest of the entries are constrained by
unitarity, i.e.
\bea
\label{ve1}
|V_{e1}| &=& \sqrt{1-|V_{e2}|^2 -|V_{e3}|^2}\\
\label{vtau3}
|V_{\tau 3}| &=& \sqrt{1-|V_{\mu 3}|^2 -|V_{e3}|^2}\\
\label{vmu1}
V_{\mu1} &=& - \frac{|V_{e2}|
  |V_{\tau3}|-|V_{e1}||V_{\mu3}||V_{e3}|e^{i\delta_{CP}}}{(1
  -|V_{e3}|^2)}\\
\label{vmu2}
V_{\mu2}
&=&  \frac{(|V_{e1}||V_{\tau3}|+|V_{e2}||V_{\mu3}||V_{e3}|e^{i\delta_{CP}})e^{i\eta}}{(1-|V_{e3}|^2)}\\
\label{vtau1}
V_{\tau1} &=&
-\frac{|V_{e2}||V_{\mu3}|+|V_{e1}||V_{\tau3}||V_{e3}|e^{i\delta_{CP}}}{(1
  -|V_{e3}|^2)}\\
\label{vtau2}
V_{\tau2} &=&
\frac{(|V_{e1}||V_{\mu3}|-|V_{e2}||V_{\tau3}||V_{e3}|e^{i\delta_{CP}})e^{i\eta}}{(1-|V_{e3}|^2)}
\eea
Using Eq.~(\ref{diagonalization}) we can express the mass matrix elements $M_{ij}$ in terms
of the $V_{PMNS}$ elements as
\begin{eqnarray}
\label{M11} M_{ee} &=& e^{2i \gamma_1} \left( |m_1|V_{e1}^2+|m_2|V_{e2}^2+|m_3|V_{e3}^2   \right)  \\
 \label{M22} M_{\mu \mu} &=& e^{2i \gamma_2} \left(  |m_1|V_{\mu1}^2+|m_2|V_{\mu2}^2+|m_3|V_{\mu3}^2   \right) \label{M22}\\
\label{M33}  M_{\tau \tau} &=& e^{2i \gamma_3} \left(   |m_1|V_{\tau1}^2+|m_2|V_{\tau2}^2+|m_3|V_{\tau3}^2  \right) \label{M33}\\
\label {M12} M_{e \mu} &=& e^{i (\gamma_1+\gamma_2)} \left(   |m_1|V_{e1}V_{\mu1}+|m_2|V_{e2}V_{\mu2}+|m_3|V_{e3}V_{\mu3}   \right) \label{M12}\\
\label{M13} M_{e \tau} &=& e^{i (\gamma_1+\gamma_3)}  \left(  |m_1|V_{e1}V_{\tau1}+|m_2|V_{e2}V_{\tau2}+|m_3|V_{e3}V_{\tau3}  \right)\label{M13}\\
\label{M23} M_{\mu \tau} &=& e^{i (\gamma_2+\gamma_3)}  \left(  |m_1|V_{\mu1}V_{\tau1}+|m_2|V_{\mu2}V_{\tau2}+|m_3|V_{\mu3}V_{\tau3}  \right)
\end{eqnarray}
Using these definitions we can now enforce that $|M_{e \mu}|=|M_{e \tau}|$
and $|M_{\mu \mu}|=|M_{\tau \tau}|$. By making a linear combination of these two
constraints we are led to an expression for the magnitude of the atmospheric mixing element
$|V_{\mu3}|$ in terms of the other neutrino sector parameters (see
section A of the Appendix for the exact expression). If we treat $|V_{e3}|$ and
the mass ratio parameter $\displaystyle r=\frac{\Delta^2_{sol}}{\Delta^2_{atm}}$ as
perturbative parameters we obtain the prediction
\bea
(|V_{\mu3}|^2)_{NH} = \frac{1}{2} - \frac{|V_{e3}|^2}{2}-r\ |V_{e1}||V_{e2}||V_{e3}|\cos{\delta_{CP}}
 \ \  +\ {\cal O}(r^2,|V_{e3}|^3)  \label{vmu3NO}
 \eea
 for the case of normal hierarchy (NH)  and
\bea
(|V_{\mu3}|^2)_{IH} = \frac{1}{2} - \frac{|V_{e3}|^2}{2}+ r\ |V_{e1}||V_{e2}||V_{e3}|\cos{\delta_{CP}}
 \ \  +\ {\cal O}(r^2,|V_{e3}|^3)  \label{vmu3IO}
\eea
for the case of inverted hierachy (IH).

This is the first  of the main results of this paper, as it represents  a general
prediction for the value of $|V_{\mu3}|$ when $\mu$-$\tau$ symmetry is
broken by phases. The NH and IH hierarchies predict slightly different
values, but the difference scales as $r |V_{e3}|$.
Due to the measured smallness of both $r$ and $|V_{e3}|$,  the
different contribution from either mass hierarchy regime is subdominant for any value of the Dirac phase
$\delta_{CP}$. Still, the scenario predicts that $|V_{\mu3}|^2$ is less than a half
 ($|V_{\mu3}|^2 <\frac{1}{2}$), with the deviation controlled  by the
term $\displaystyle \frac{|V_{e3}|^2}{2}$. A  distinction between NH and IH in the predicted
value of $|V_{\mu3}|$ is a very interesting result, although challenging to test experimentally.

The second original prediction that we obtain comes from the two constraints $|M_{e \mu}|=|M_{e \tau}|$
and $|M_{\mu \mu}|=|M_{\tau \tau}|$ and it corresponds to is a sum rule equation relating all the
physical parameters from the neutrino sector. Again treating
$|V_{e3}|$ 
as a perturbative parameter, we can extract  a relatively simple approximate
sum rule given by
\bea
\cos(\delta_{CP})\ K_1 + \sin{(\delta_{CP})}\ K_2 \ =\  0   \  \ +\ \ {\cal O}(|V_{e3}|^2)  \label{sumruletext}.
\eea
where $K_1$  and $K_2$ are
\bea
K_1&=& |m_2|^2|V_{e2}|^2-|m_1|^2  (1-|V_{e2}|^2) + (1 -2  |V_{e2}|^2)
|m_1||m_2| \cos(2\eta)\non\\
 &&\ +|m_1||m_3|\cos(2\xi)-|m_2||m_3|\cos(2\eta-2\xi)\\
K_2&=& |m_1||m_2|\sin(2\eta)+|m_1||m_3|\sin(2\xi)+|m_2||m_3|\sin(2\eta-2\xi).
\eea
Note that when $\delta_{CP} = \pm \pi/2$ the sum rule becomes an exact
relation to all orders in $|V_{e3}|$  and is given by
\bea
\label{sumruledelpio2}
|m_1||m_2|\sin(2\eta)+|m_1||m_3|\sin(2\xi)+|m_2||m_3|\sin(2\eta-2\xi)\
 =\ 0  \ \ \ \ \ \  ({\rm for}\ \   \delta_{CP} =\pm
 \pi/2). \ \ \ \ \ \
 \eea
 which shows that the two Majorana phases $\eta$ and $\xi$ are explicitly linked to the neutrino
 masses in a very simple way. Details of the exact analytical
 expressions will be given in the appendix.

 Another interesting limit associated to the approximate sum rule is
 when the lightest neutrino mass is zero. In the {\it Normal hierarchy} case,
 this limit is obtained by setting $m_1 \to 0$ and $m_2^2 \to \Delta^2_{sol}$  and
 $m_3^2 \to \Delta^2_{atm}$, leading to
 \bea
 \cos{(2 \eta-2\xi+\delta_{CP} )} - \cos{(\delta_{CP})}\  \sqrt{r}
 |V_{e2}|^2\ =\ 0   \ \  +\ \  {\cal O}(|V_{e3}|^2)   \ \ \ \ \ ({\rm
   for \ \ m_1} \to0) \ .
 \eea
For the {\it Inverted hierarchy} case,
we set the limits $m_3\to 0$ and $m_2^2\to (\Delta^2_{sol}+\Delta^2_{atm})$  and
$m_1^2\to \Delta^2_{atm}$ leading to
\bea
\tan{(\delta_{CP})}  &=& (1-2|V_{e2}|^2)  \tan{(\eta)}
\ \ \   \left(1\   + \  {\cal    O}(|V_{e3}|^2) \right)
\ \ \ \ \ ({\rm for}\ \  m_3 \to 0)
 \eea
 These two approximate relations show that when the lightest neutrino mass is
 zero (or very small) the Majorana phases $\eta$ and $\xi$ and the Dirac CP
 phase $\delta_{CP}$ obey very simple approximate sum-rule relations
 when the exact $\mu-\tau$ symmetry is broken by phases.


 Furthermore the dependence of the phase $\theta$ (which breaks the
 $\mu-\tau$ {\it permutation} symmetry) can be written explicitly in
 terms of the other physical parameters. Using
 Eq.~(\ref{phasebrokenmutau}) with  $|M_{e \mu}|=|M_{e \tau}|$
and $|M_{\mu \mu}|=|M_{\tau \tau}|$ we write
 \bea
M_{\mu \mu}&=& e^{2i (\theta-2\sigma)} M_{\tau \tau}\\
M_{e \mu}&=&e^{2 i(\theta-\sigma)}M_{e \tau}.
\eea
With this we can formally obtain the relation (which we will refer to
as the ``{\it $\theta$-equation}'') between the phase
breaking parameter $\theta$ and the rest of neutrino sector parameters as
\bea
e^{2 i\theta} =\left(\frac{M_{\tau\tau}M^2_{e\mu}}{M_{\mu\mu}M^2_{e\tau}}\right)
\label{thetatext}
\eea
where that particular combination of $M_{ij}$'s depends only on physical parameters of
the neutrino sector as the unphysical phases $\gamma_i$ happen to
cancel out (see Eqs.~(\ref{M22})-(\ref{M13}) as well as the Appendix for more
details).
We will use this relation in the next section.


Finally, using the previous constraints, it is also possible to extract an analytical expression for the mixing angle
$|V_{e3}|$ in terms of the neutrino mass matrix elements (as defined
in Eq.(\ref{phasebrokenmutau})) as well as in terms of $\Delta^2_{sol}$,
$\Delta^2_{atm}$ ,$|V_{e2}|$ and $\delta_{CP}$. We find
\begin{eqnarray}
  |V_{e3}|&\simeq&\pm
  \frac{2     N_1  }{\Delta_{sol}^2\Delta_{atm}^2|V_{e2}|\sqrt{1-|V_{e2}|^2}}
  \sin(\theta)
\end{eqnarray}
where $N_1$ depends only on mass matrix elements as
\bea
N_1&=&\sqrt{\Big(Im(a^*b)\Big)^2+\frac{4}{\left(\Delta^2_{atm}\right)^2}
  \Big[(|b|^2-|a|^2) Re(c) + d \Big(Re(a b^*)+|b|^2\cos{(\theta)}\Big)\Big]^2 }
\eea
with $a=m_{11}m^*_{12}+m_{12}m^*_{23} $,\  $b=m_{12}m_{22}^*$,\ $c=m_{22}m^*_{23}$\ and\ $d=|m_{12}|^2$.
The main message from this expression is how indeed we can recover
exact $\mu$-$\tau$ {\it permutation }symmetry
by setting $\theta=0$, in which case $|V_{e3}|$ vanishes and we
also  have $|V_{\mu 3}|= |V_{\tau 3}|=\frac{1}{2}$. \footnote{ Note also that
within exact $\mu$-$\tau$ {\it permutation} symmetry, the unphysical phases
$\gamma_i$ appearing in the diagonalization of the neutrino mass
matrix (see Eq.~(\ref{Umatrix})) are such that $\gamma_1=0$ and $\gamma_2=\gamma_3$.}

The expression for $|V_{e3}|$ also shows that in the different limit of $\mu$-$\tau$ {\it reflection
symmetry} (which in this context is a particular case of {\it
  phase-broken} $\mu$-$\tau$), $|V_{e3}|$ does not necessarily
vanish. Indeed in that limit the matrix elements $m_{11}$, $m_{12}$,
$m_{23}$ and $m_{22}$ are all real and the smallness of $|V_{e3}|$ can
either be caused by the smallness of $N_1$, or due to a small value
of the phase  $\theta$.

\subsection{$\mu$-$\tau$ reflection symmetry (an example of phase  broken $\mu$-$\tau$)}
\label{sec:mu-tau-reflection}
Within $\mu$-$\tau$ {\it reflection symmetry} \cite{Harrison, Grimus,Valle2, Xing, ReflectionOther}, the constraints imposed on the
neutrino mass matrix are such that $M_{ee}$ and $M_{\mu\tau}$ are real
and $M_{e\mu}=M^{*}_{e\tau}$ and  $M_{\mu\mu}=M^*_{\tau\tau}$.
The mass matrix can be written using the parametrization we have used for
phase-broken $\mu$-$\tau$ (see Eq.~(\ref{phasebrokenmutau})) as
\begin{eqnarray}
    M_{\nu}\equiv
  \left(\begin{array}{ccc} M_{ee} & M_{e \mu} &  M^*_{e \mu} \\
  M_{e \mu} & M_{\mu \mu} & M_{\mu \tau} \\
  M^*_{e \mu} & M_{\mu \tau} & M^*_{\mu \mu} \end{array}\right)=\
  P_\sigma\ \left(\begin{array}{ccc} m_{11}\ \ \ &
    m_{12} e^{i\theta} & \ \  m_{12} e^{-i\theta} \\
  m_{12} e^{i\theta} & m_{22} e^{i\theta} & m_{23}  \ \ \\
  \ \  m_{12} e^{-i\theta} & m_{23} \ \  &\ \  m_{22} e^{-i\theta} \end{array}\right)\ P_\sigma,
\end{eqnarray}
where again $P_\sigma = diag(1, e^{-i\sigma}, e^{i\sigma})$ but now
all the entries $m_{ij}$ are real and all the phase information is
parametrized by the single phase breaking parameter $\theta$ (the
phase $\sigma$ remains unphysical as it is included in the diagonal
phase matrix $P_{\sigma}$).

This symmetry clearly corresponds to a special case of phase-broken
$\mu$-$\tau$ symmetry (with the phase $\theta$) but the extra
reality constraints will further limit the allowed parameter space. In
particular it is well known that under  $\mu$-$\tau$ {\it reflection
  symmetry} the physical phases are fixed, i.e. the Dirac CP phase is
such that $\delta_{CP}=\pm \pi/2$, and the Majorana phases are such
that  $\eta=0$ or $ \pi/2$ and $\xi=0$ or $\pi/2$.
It is also known that the neutrino mixing angles are such that
\bea
|V_{\mu3}|^2= |V_{\tau3}|^2 = \frac{1}{2}(1-|V_{e3}|^2)
\label{mutaureflectionvmu3}
\eea
and that there are no constraints on the physical neutrino masses.

These results are perfectly consistent with the general predictions
of  phase  broken $\mu$-$\tau$ (see Equations (\ref{vmu3NO}),
(\ref{vmu3IO}) and (\ref{sumruletext})) where we see
that setting  $\delta_{CP}=\pm \pi/2$ leads an agreement with
Eq.~(\ref{mutaureflectionvmu3}) and setting $\eta=0$ or $\pi/2$,
$\xi=0$ or $\pi/2$ eliminates the sum-rule constraint linking neutrino masses and
mixings shown in Eq.~(\ref{sumruletext}).

Under our parametrization, however, we would like to point out that we can
link in an exact and straightforward way the phase breaking parameter
$\theta$ (a phase parameter from the neutrino mass matrix)  to the
physical observables of the neutrino sector.

In particular we can exploit the {\it ``$\theta$-equation''} given by
Eq.~(\ref{thetatext}) in the special limit of exact $\mu$-$\tau$
reflection symmetry (and choosing the case $\delta_{CP} =
-\pi/2$) \footnote{For the case  $\delta_{CP} = +\pi/2$ the relation changes by an overall sign, and so the associated curves would be
  the mirror image of the curves shown in Figure 1.} and find the
exact relation
\bea
\tan(\theta) &=& - \frac{2|V_{e1}||V_{e2}||V_{e3}| (m_2-|m_1|)(|m_1|+m_3)(m_2+m_3)}{C_0+C_1|V_{e3}|^2+C_2|V_{e3}|^4+C_3|V_{e3}|^6}
\label{thetasol}
\eea
where we have defined
\bea
\label{C0} C_0 &=& (m_2-|m_1|)^2|V_{e2}|^2(1-|V_{e2}|^2)\left[(m_2+m_3)-(m_2-|m_1|)|V_{e2}|^2\right] \\
\label{C2}C_2 &=& (|m_1|+m_3)^2\left[|m_1|+m_2+2m_3+3(m_2-|m_1|)|V_{e2}|^2\right]   \\
\label{C3}C_3 &=& -(|m_1|+m_3)^3
\eea
and,
\bea
\label{C1}C_1 &=&
-(|m_1|+m_3)\left[(|m_1|+m_3)(m_2+m_3)+(m_2-|m_1|)(2|m_1|-m_2+m_3)|V_{e2}|^2\right]
\nonumber \\
&&-3(|m_1|+m_3)(m_2-|m_1|)^2|V_{e2}|^4.
\eea
Note that the mass parameters are all real but we have allowed $m_2$ and
$m_3$ to be either positive or negative (due to the different possible
values of the Majorana phases,  $\eta=0$ or $\pi/2$ and $\xi=0$ or
$\pi/2$). We thus have $m_2= \pm |m_2|$ and $m_3=\pm|m_3|$ and
therefore there are 4 distinct possible solutions within
Eq.(\ref{thetasol}). The denominator is written as a polynomial in
$|V_{e3}|$ to showcase that  the coefficients of $|V_{e3}|^4$ and $|V_{e3}|^6$ can
be neglected numerically due to the smallness of the experimental
value for $|V_{e3}|$.  If one neglects these terms and further expands the expressions using
  $\displaystyle r=\frac{\Delta^2_{sol}}{\Delta^2_{atm}}$ as a small
parameter, we obtain a quadratic expression in $|V_{e3}|$ leading to a
group of very simple possible expressions for the mixing angle $|V_{e3}|$ as a function of the
physical neutrino masses, the mixing angle $|V_{e2}|$ and the phase breaking parameter $\theta$ as
\bea
|V_{e3}| \simeq - \frac{(m_2-|m_1|)}{(|m_1|+m_3)}
\frac{\sqrt{1-|V_{e2}|^2}\ |V_{e2}|}{ \left(1 + \frac{(m_2-|m_1|)}{(m_2+m_3)} |V_{e2}|^2\right)}  \tan(\frac{\theta}{2})
\label{thetasolapprox1}
\eea
or
\begin{eqnarray}
|V_{e3}| \simeq  \frac{(m_2-|m_1|)}{(|m_1|+m_3)}
\frac{\sqrt{1-|V_{e2}|^2}\ |V_{e2}|}{ \left(1 +
 \frac{(m_2-|m_1|)}{(m_2+m_3)} |V_{e2}|^2\right)}
\cot(\frac{\theta}{2})
\label{thetasolapprox2}
\end{eqnarray}
where again $m_2= \pm |m_2|$ and $m_3=\pm|m_3|$, the signs depending on the
values of the Majorana phases.\\

The exact expression in  Eq.~(\ref{thetasol}) along with the approximative results  (\ref{thetasolapprox1}) and
(\ref{thetasolapprox2}) represent, to our knowledge, a new
 contribution to  the well established scenario of  $\mu$-$\tau$ {\it reflection
   symmetry}. 

\begin{figure}[t]
  \center
  \includegraphics[height=7cm,width=8cm]{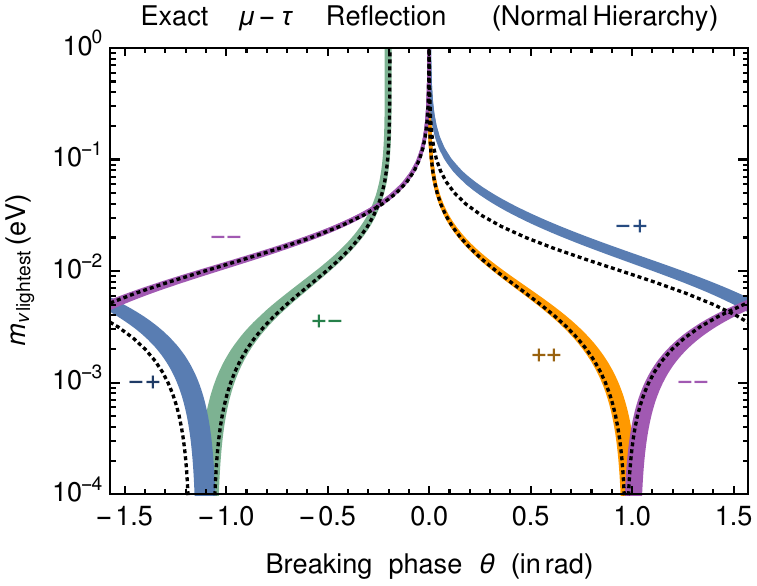}\hspace{.1cm}
 \includegraphics[height=7cm,width=8cm]{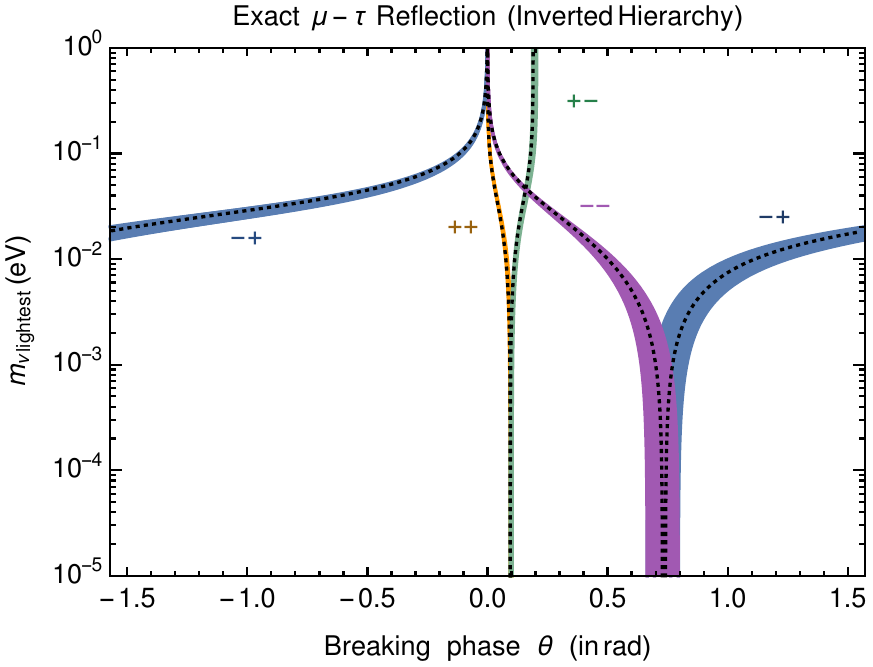}
 \caption{Lightest neutrino mass as a function
   of the phase $\theta$ (which breaks $\mu-\tau$ {\it permutation}
   symmetry) in the case of {\it exact}  $\mu-\tau$ {\it reflection}
   symmetry (an example of phase-broken $\mu-\tau$ {\it
     permutation}). Normal (left) and Inverted (right) Hierarchies are
   considered and the ``$++$,''
   ``$+-$'', ``$-+$'' and  ``$- -$'' curves correspond to the
   different possible values of the Majorana phases ($\eta$=0; $\xi$=0),
   ($\eta$=0; $\xi= \pi/2)$, ($\eta= \pi/2$; $\xi$=0) or
   ($\eta=\pi/2$; $\xi=\pi/2$).
   The curves represent points where the mixing angles $|V_{e3}|$
   and $|V_{e2}|$ and the mass differences $\Delta^2_{sol}$ and
   $\Delta^2_{atm}$ lie within their 1-$\sigma$
   experimental range.  Dotted curves represent the simple analytical
   approximations from Eqs.~(\ref{thetasolapprox1}) and
   (\ref{thetasolapprox2}) where we have used  only the central values of
   experimental data.  For Normal Hierachy and very light mass, the phase
   $\theta$  can only be $\theta_0 \simeq 1  (mod\ \pi)$ or $\theta_0 \simeq -1.1  (mod\ \pi)$, whereas for
   Inverted Hierarchy we have $\theta_0 \simeq 0.1  (mod\ \pi)$ or $\theta_0 \simeq 0.7  (mod\ \pi)$, (plots are symmetric under $\theta \to \theta -\pi$).}
 \label{fig:mutaureflection}
 \end{figure}


We show graphically these dependencies in figure
\ref{fig:mutaureflection}  in which we plot  the
lightest neutrino mass as a function of the phase parameter
$\theta$ for the cases of Normal hierarchy (left panel) and Inverted
hierarchy (right panel). The
mixing angles $|V_{e3}|$ and $|V_{e2}|$ have been
allowed to range in their $1-\sigma$ experimental range and we have
also fixed the value of the solar and atmospheric neutrino mass
differences to their experimental values, so that we obtain bands of
allowed points. We have used the exact expression from
Eq.(\ref{thetasol}),  but have also included the approximate expressions from
Eqs.~(\ref{thetasolapprox1})  and (\ref{thetasolapprox2}) in order to
track their validity (dashed curves, obtained using only the central values
of the mixing angles).
We clearly see that once $|V_{e3}|$, $|V_{e2}|$ and the experimental
neutrino mass differences are fixed, the value of
$\theta$ determines very precisely the possible lightest neutrino mass. In particular when
the lightest mass tends to zero only two values of  $\tan(\theta)$ are
allowed within the Normal Hierarchy case ($\tan(\theta)\simeq \pm 1.5$)
and only two other values are possible for the Inverted Hierarchy
case, ($\tan(\theta) \simeq 0.1$ or $\tan(\theta)\simeq 0.9$). Because
the function $\tan(\theta)$ has a periodicity of $\pi$ we show only
values of $\theta$ between $-\pi/2$ and $\pi/2$.

Note that no other lagrangian parameter enters in these relations showcasing the
importance of the phase breaking parameter $\theta$ within  {\it
  exact}  $\mu$-$\tau$ {\it reflection  symmetry}.

\subsection{Phase broken $\mu$-$\tau$ with $\delta_{CP}=-\pi/2$}
\label{sec:delmpihalf}

In this section we are going to study the particular case where we fix
$\delta_{CP}=-\pi/2$, in part guided by the exact analytical results
from the previous section\footnote{
Note that we are allowing the Majorana phases $\eta$ and $\xi$ to have any value and
thus this case does not correspond to  $\mu$-$\tau$ {\it reflection
   symmetry} in which $\eta=0$ or $\pi/2$ and
$\xi=0$ or $\pi/2 $.} and also in part guided by the global fit results from
neutrino phenomenology which place $\delta_{CP}$  somewhere around the
third quadrant. On the other hand, and unlike the previous section, the
Majorana phases can take any value (as long as the model constraints
are verified).
It turns out that in this limit the analytical results
are still quite simple, giving us a further robust understanding
of the  features of phase broken $\mu-\tau$ symmetry, without having
to rely on numerical solutions.

\begin{figure}[t]
  \center
  \includegraphics[height=8cm,width=8cm]{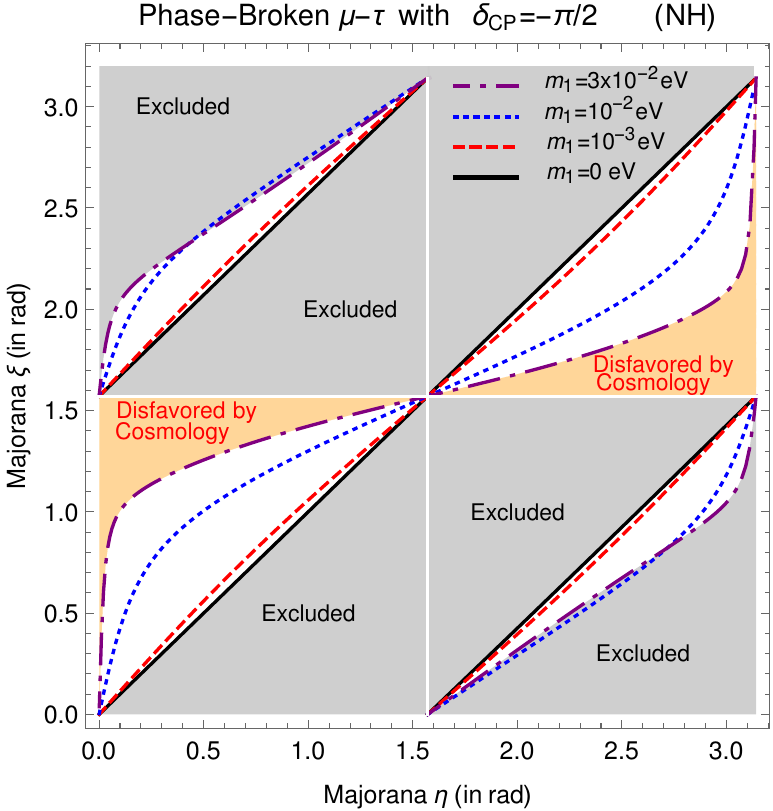}
  \includegraphics[height=8cm,width=8cm]{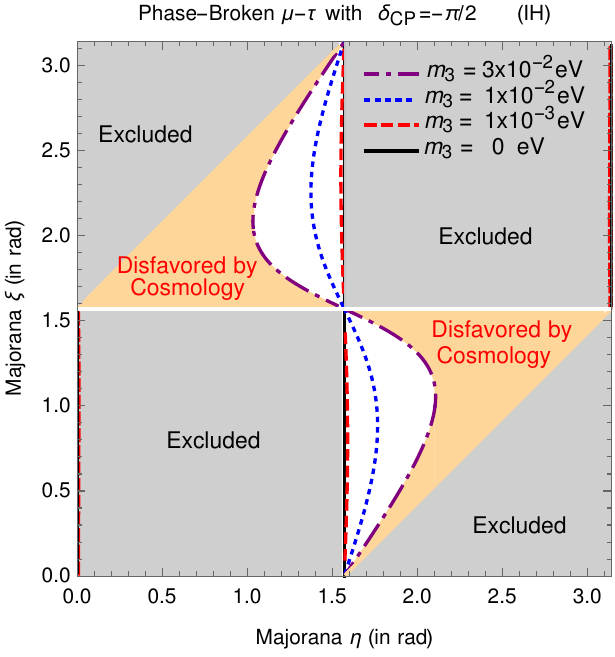}
  \caption{
Contours of the lightest neutrino mass as a function
of the Majorana phases $\eta$ and $\xi$ within phase-broken $\mu-\tau$  in the limit where the Dirac CP phase in the neutrino
sector is $\delta_{CP}=- \pi/2$. The excluded regions on both panels
(left for Normal and right for Inverted Hierarchies)
represent  values of $\eta$ and $\xi$ inconsistent with the sum-rule
in eq.~(\ref{sumrulepio2}). Note that because the sum-rule depends on $(2\eta)$ and
$(2 \xi)$ it is enough to vary their values between $0$ and $\pi$.
The regions marked ``Disfavored by Cosmology'' are
regions where the lightest neutrino mass value is too large to evade
cosmological constraints \cite{cosmology}. 
Also,  for the specific values of $0$ and $\pi/2$ for the Majorana
phases, the constraint on neutrino masses disappears.
    }
\label{fig:etaxi}
\end{figure}

We start with the two constraints $|M_{e \mu}|=|M_{e \tau}|$
and $|M_{\mu \mu}|=|M_{\tau \tau}|$ along with $\delta_{CP}=-\pi/2$.
This leads to the equality of the moduli of the elements of the
second row and the third row of $V_{PMNS}$, i.e $|V_{\mu i}|=|V_{\tau
  i}|$ .\footnote{Note that exact {\it permutation} and {\it
    reflection} $\mu-\tau$ symmetries do also lead to the equality
  $|V_{\mu i}|=|V_{\tau i}|$. }
More specifically we cast the consequences of this particular limit of
phase broken $\mu-\tau$ with two simple and exact
relations, namely a prediction for $|V_{\mu3}|$
\bea
|V_{\mu3}| = \frac{\sqrt{1-|V_{e3}|^2}}{\sqrt{2}}
\eea
and the exact sum rule (already given in Eq.~(\ref{sumruledelpio2}))
\bea
\label{sumrulepio2}
|m_1||m_2|\sin(2\eta)+|m_1||m_3|\sin(2\xi)+|m_2||m_3|\sin(2\eta-2\xi)\  =\ 0
\eea
in which the majorana phases $\eta$ and $\xi$ are constrained along with
the neutrino masses in a very simple way.
\begin{figure}[t]
  \center
  \includegraphics[height=7cm,width=8cm]{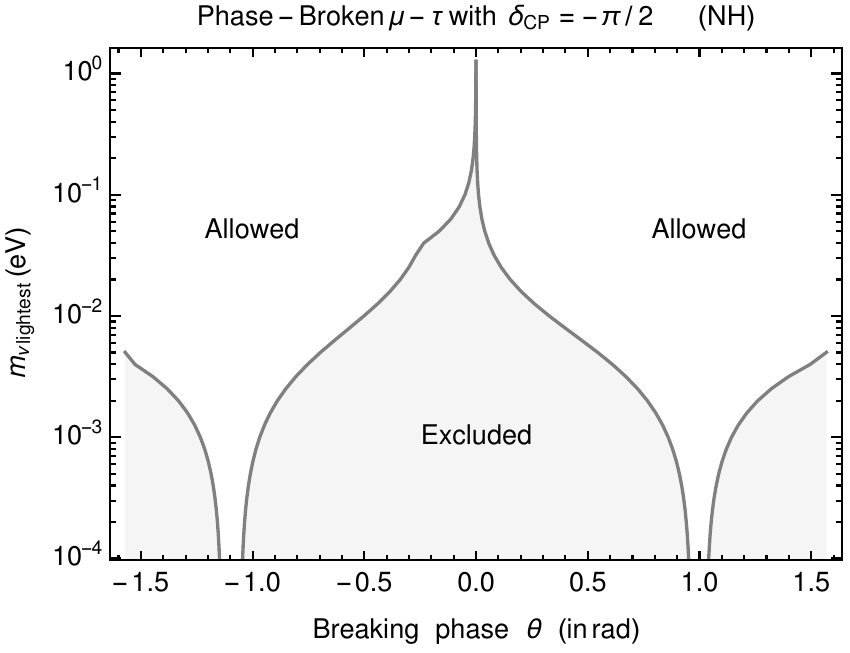}\hspace{.1cm}
 \includegraphics[height=7cm,width=8cm]{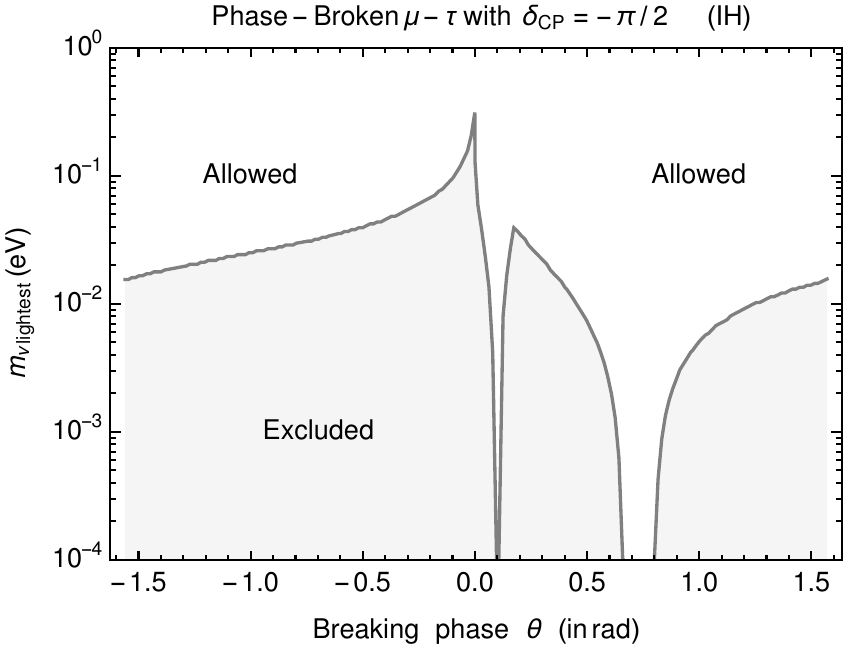}
 \caption{Lightest neutrino mass as a function
   of the phase $\theta$ (which breaks $\mu-\tau$ {\it permutation}
   symmetry) in the limit where the Dirac CP phase in the neutrino
   sector is $\delta_{CP}=- \pi/2$, with the Majorana phases $\eta$
   and $\xi$ not fixed.  The allowed regions represent points in which
   the mixing angles $|V_{e3}|$ and $|V_{e2}|$ and the mass
   differences $\Delta^2_{sol}$ and   $\Delta^2_{atm}$ are constrained
   to lie  within their 1-$\sigma$ experimental range.  When the
   lightest neutrino mass is very small, for Normal Hierachy  the
   phase    $\theta$  can only be $\theta_0 \simeq 1  (mod\ \pi)$ or
   $\theta_0 \simeq -1.1  (mod\ \pi)$, whereas for  Inverted Hierarchy
   we have $\theta_0 \simeq 0.1  (mod\ \pi)$ or $\theta_0 \simeq 0.7
   (mod\ \pi)$, (plots are symmetric under $\theta \to \theta -\pi$).}
\label{fig:mutaupio2}
\end{figure}
Note that for the specific values $0$ or $\pi/2$ for these phases,
all the terms of the sum rule vanish and we lose any
constraints on the neutrino masses. When the Majorana
phases take these values, we obtain the exact $\mu-\tau$ {\it
  reflection} symmetry limit, which can be understood as a particular
case of phase broken $\mu-\tau$ symmetry when $\delta_{CP}=\pm
\pi/2$. The predictions coming from the sum-rule in
eq.~(\ref{sumrulepio2}) are displayed in figure  \ref{fig:etaxi}, and
we observe that vast regions of the $(\eta-\xi)$ parameter space are
inconsistent with the sum-rule. Cosmology constraints \cite{cosmology}
also affect importantly the allowed parameter space region as shown in the figure.

Performing a similar scan as in figure  \ref{fig:mutaureflection}, we
study in figure \ref{fig:mutaupio2} the relation between the
lightest neutrino mass in terms of the phase parameter
$\theta$ for the cases of Normal hierarchy (left panel) and Inverted
hierarchy (right panel).
The mixing angles $|V_{e3}|$ and $|V_{e2}|$ have been
allowed to range in their $1-\sigma$ experimental range and we have
fixed the value of the solar and atmospheric neutrino mass
differences to their central experimental values.
In this case we observe that even though $\delta_{CP}$ remains fixed
at $=-\pi/2$, the relaxation of the Majorana phases does open an
allowed parameter space region that was closed in the $\mu-\tau$ reflection
symmetry limit studied in the previous section. We observe two
regions, allowed and excluded and we realize that the border of the
regions are actually the curves for fixed values of the Majorana
phases with $\eta= 0$ or $\pi/2$ and $\xi = 0$ or $\pi/2$, i.e. the
limiting curves correspond to $\mu-\tau$ reflection symmetry points,
shown in figure  \ref{fig:mutaureflection}.

\begin{figure}[t]
  \center
  \includegraphics[height=8cm,width=12cm]{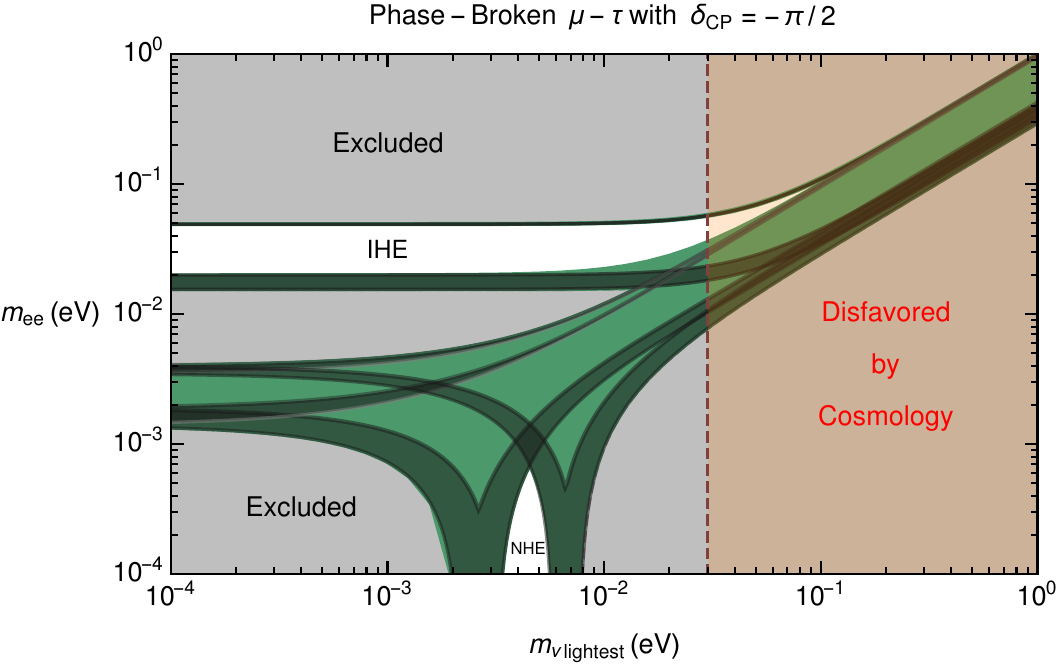}
 \caption{Allowed values (green/darker green) of the effective Majorana mass $m_{ee}$ in
   terms of the lightest neutrino mass in phase-broken $\mu$-$\tau$
   in the limit where the Dirac CP phase in the neutrino  sector is
   fixed to $\delta_{CP}=- \pi/2$, with the Majorana phases $\eta$ and
   $\xi$ allowed to range within the model theoretical constraints.
   The dark green regions are points with $\eta$ and
   $\xi$ fixed to $\pi/2$ or $0$, representing the allowed regions
   within exact  $\mu$-$\tau$ {\it reflection symmetry}.
   The white regions marked ``IHE'' and ``NHE''
   are excluded due to the phase-broken $\mu$-$\tau$ constraints (in
   the Inverted and Normal Hierarchy regimes),   whereas the gray
   regions marked ``Excluded'' violate current experimental
   constraints on neutrino masses and mixings.
}
\label{fig:mee}
 \end{figure}

Another interesting phenomenological aspect to consider is the
effective Majorana neutrino mass which is given by
\begin{eqnarray}
|m_{ee}|= \Big|  |m_1| |V_{e1}|^2+|m_2||V_{e2}|^2e^{2i\eta}+|m_3||V_{e3}|^2e^{2i(\xi-\delta_{CP})}\Big|
\end{eqnarray}
New generation and near future experiments \cite{comingexperiments}
will be sensitive to an important region of the allowed parameter
space (in the Inverted Hierarchy) so that predictions from phase
broken $\mu-\tau$ symmetry could be directly probed.

In figure \ref{fig:mee} we show in  green/darker green the allowed
values of the effective Majorana mass $m_{ee}$ in  terms of the
lightest neutrino mass in phase-broken $\mu$-$\tau$  for
$\delta_{CP}=- \pi/2$, and with the Majorana phases $\eta$ and  $\xi$
allowed to range within the phase-broken $\mu-\tau$ constraints.  The
dark green regions are points where $\eta$ and  $\xi$ are fixed to
$\pi/2$ or $0$, and so represent the allowed regions  for exact
$\mu$-$\tau$ {\it reflection symmetry}.   The upper and lower
gray areas are points excluded by current experimental constraints on
neutrino masses and  mixings, whereas the brown area on the right
side represents  disfavored  points from cosmology \cite{cosmology}.
We observe that broken $\mu-\tau$ scenario puts constraints on two regions
of parameter space, shown in white and marked ``IHE'' (Inverted
Hierarchy Exclusion) and ``NHE'' (Normal Hierarchy Exclusion)
representing points excluded by the phase-broken $\mu$-$\tau$
constraints (in the inverted and normal hierarchy regimes).

\subsection{Phase broken $\mu$-$\tau$ with specific values for $\delta_{CP}$.}
\label{sec:delother}

\begin{figure}[t]
  \center
  \includegraphics[height=7cm,width=8cm]{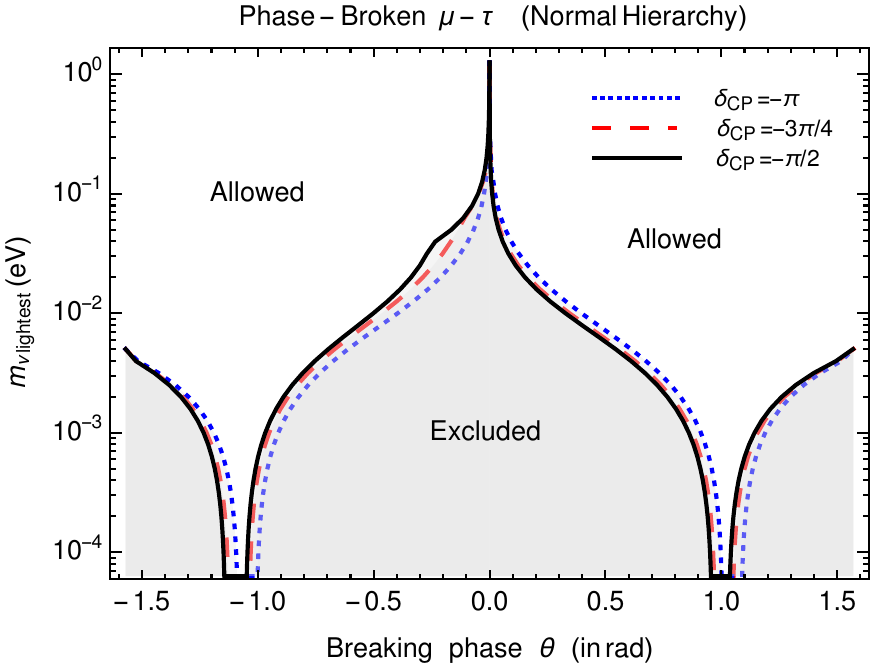}\hspace{.1cm}
  \includegraphics[height=7cm,width=8cm]{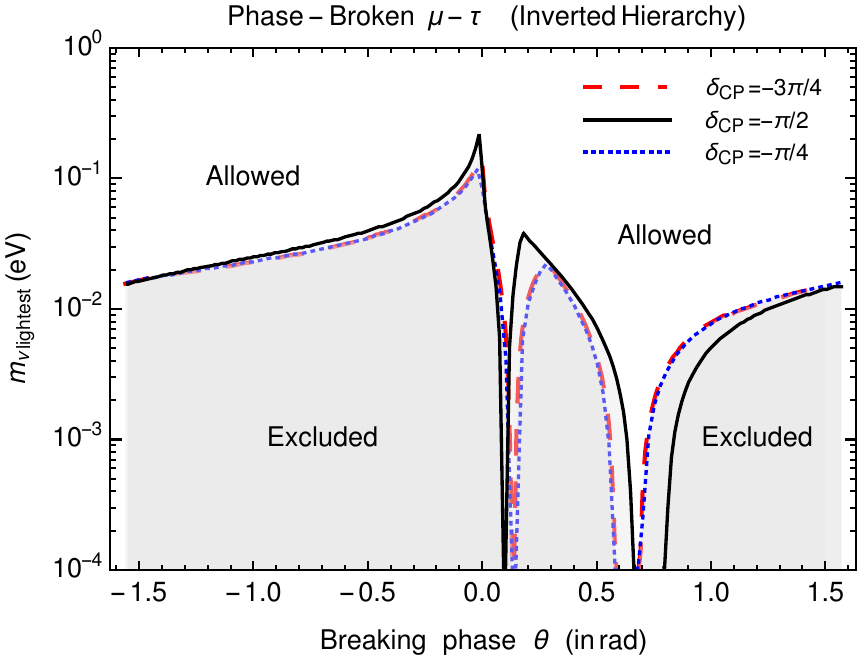}\hspace{.1cm}
 \caption{Lightest neutrino mass as a function
   of the phase $\theta$ (which breaks $\mu-\tau$ {\it permutation}
   symmetry) when the Dirac CP phase in the neutrino
   sector takes the values $\delta_{CP}=(- \pi/2)$, $(-3\pi/4)$, and
   $(-\pi)$ for Normal Hierarchy  and $\delta_{CP}=(- \pi/4)$, $(-\pi/2)$, and
   $(-3\pi/4)$ for Inverted Hierarchy. The Majorana phases $\eta$ and
   $\xi$ are not fixed.  The allowed regions represent points in which
   the mixing angles $|V_{e3}|$ and $|V_{e2}|$ and the mass
   differences $\Delta^2_{sol}$ and   $\Delta^2_{atm}$
   are constrained to lie  within their 1-$\sigma$ experimental global
   fit range.
 }
\label{fig:mutaureflectiongeneral}
\end{figure}

We finish our overview of the phase broken $\mu-\tau$ symmetry scenario by
relaxing the fixed value of $\delta_{CP}$ and cover the range of values
prefered by the global fits from neutrino data.
We would like to explore here how sensitive are the previous results
to relatively small changes in the value of $\delta_{CP}$. Current
global fits from experimental data on masses and mixings in the
neutrino sector more or less point towards a value of $\delta_{CP}$
roughly located within the range $-\pi \lesssim \delta_{CP} \lesssim -\pi/2 $ for
Normal Hierarchy, and $-3\pi/4 \lesssim \delta_{CP} \lesssim -\pi/4$
for Inverted Hierarchy.
We thus wanted to see how the parameter space is constrained within
these windows of parameter space.

In figure \ref{fig:mutaureflectiongeneral} we show again the
dependance of the lightest neutrino mass with respect to the phase-breaking
parameter $\theta$ (which breaks the $\mu-\tau$ permutation symmetry)
in both the Normal Hierarchy (NH) and Inverted Hierarchy (IH) frameworks.
In each case we fix the value of the Dirac phase to $\delta_{CP} =
-\pi$, $-3\pi/4$ and -$\pi/2$ (for NH) and  $\delta_{CP} = -3\pi/4$,
$-\pi/2$ and -$\pi/4$ (for IH).
The values for $\delta_{CP} = -\pi/2$ are the same as in figure \ref{fig:mutaupio2}
and the associated points are shown as a continuous black
curves. For $\delta_{CP} = -3\pi/4$ we plot the results as red/dashed
curves and for  $\delta_{CP} = -\pi$ (NH) and $-\pi/4$ (IH) we show
the results in blue/dotted curves.

The regions below the curves are excluded by the phase broken $\mu-\tau$
constraints whereas the white regions represent points that are
allowed by all theoretical and experimental constraints. We can see
that there is not a strong dependence on the value
of  $\delta_{CP}$, at least within this window, so that the results
from the previous section are quite typical within this extended
parameter space.

We still observe that for very light neutrino mass, the phase breaking
phase $\theta$ is extremely constrained. In the NH case its values must
be $\theta_0 \simeq 1$  $(mod\ \pi)$ or $\theta_0 \simeq -1.1$
$(mod\ \pi)$, with essentially no dependance on the value of
$\delta_{CP}$ or the Majorana phases $\eta$ and $\xi$.
For  Inverted Hierarchy we have $\theta_0 \simeq 0.1$  $(mod\ \pi)$ or
$\theta_0 \simeq 0.7$-$0.8$  $(mod\ \pi)$, this time with a mild
dependance on $\delta_{CP}$, but no dependance on the Majorana phases
$\eta$ and $\xi$.

We can prove analytically these features by analyzing the phase broken
$\mu-\tau$ constraints in the limits $m_1=0$ (NH) or $m_3=0$ (IH),
and performing an expansion in $|V_{e3}|$ and the mass ratio parameter
$\displaystyle r=\frac{\Delta^2_{sol}}{\Delta^2_{atm}}$. We find, to
lowest order,  the analytical expression
\bea
\tan{\theta}\simeq \pm \frac{2 \sqrt{r}\sqrt{1-|V_{e2}|^2} |V_{e2}|
  |V_{e3}|}{r |V_{e2}|^2 (1-|V_{e2}|^2)-|V_{e3}|^2}
\eea
for NH, and for IH we obtain two possible values of the phase breaking
parameter
\begin{eqnarray}
\tan(\theta)&\simeq& -\frac{r|V_{e1}||V_{e2}|}{|V_{e3}|}\sin(\delta)
\end{eqnarray}
or
\begin{eqnarray}
\tan(\theta)&\simeq&
-\frac{\left[1-4|V_{e1}|^2|V_{e2}|^2\cos^2(\delta)\right]|V_{e3}|\sin(\delta)}{|V_{e1}||V_{e2}|\left[1-2|
   V_{e2}|^2\right]}
\end{eqnarray}
It is easy to check that using the experimental values for  $r$, $|V_{e2}|$
and  $|V_{e3}|$, and including the appropriate values of
$\delta_{CP}$, one reproduces the values of $\theta$ located at the ``throats'' of
the numerical curves of Figure \ref{fig:mutaureflectiongeneral}.

We note however that these expressions are completely general
predictions of the phase-broken $\mu-\tau$ paradigm, for any value of
$\delta_{CP}$ as long as we are in the limit of a massless (or vey
light) lightest neutrino. It is quite non-trivial that the dependance
on the Majorana phases disappears in the limit of vanishing mass for
the lightest neutrino.

\section{Conclusions}
\label{sec:conclusion}
In this paper we have considered deviations from the usual
$\mu$-$\tau$ symmetry framework by adding general phases to a
$\mu$-$\tau$ symmetric neutrino mass matrix. We called the obtained
framework {\it phase-broken} $\mu$-$\tau$ symmetry and studied its
general predictions.

As a general result of the setup, the atmospheric mixing element $|V_{\mu3}|^2$ is predicted to be less
than a half ($|V_{\mu3}|^2 <\frac{1}{2}$), with the deviation set by
$|V_{e3}|^2/2$.  Also a different value for $|V_{\mu3}|$  is predicted
for the cases of NH and IH mass orderings, although the difference is
numerically very small and thus will be quite challenging to test
experimentally.


One example of our framework turns out to be a very well studied
version of $\mu$-$\tau$ symmetry, namely the $\mu$-$\tau$ reflection
symmetry. Indeed the  $\mu$-$\tau$ reflection symmetry can be
understood as a phase-broken $\mu$-$\tau$ permutation symmetry, with
additional constraints on the mass matrix elements (namely that the
only new phase lies in the phase-breaking phase $\theta$).
With this point of view we were able to extract new relations
associated to $\mu$-$\tau$ reflection symmetry (to our knowledge) and in
particular we obtained analytical expressions of the mixing angle $|V_{e3}| $ in terms of masses and
mixing, involving only the phase breaking parameter $\theta$ out of the
Lagrangian parameters.

We then studied relaxations of the $\mu$-$\tau$ reflection symmetry
within the point of view of {\it phase-broken} $\mu$-$\tau$ symmetry,
by scanning the paramater space staying in regions where the value of
the Dirac CP phase $\delta_{CP}$ is close to $-\pi/2$.

As a general feature in this region of parameter space we observe that
in order to allow for a very light neutrino mass, the model
requires the phase breaking parameter to have precise
values (two different allowed values for Normal Hierarchy and two
other for Inverted Hierarchy). Heavier lightest neutrino masses will relax the constraint,
but the model still showcases a direct sensitivity to a single
lagrangian parameter, the phase $\theta$.

\section{Acknowledgments}
E.I. Lashin and N. Chamoun acknowledge support from ICTP-Associate
program. N.C. acknowledges support of the Alexander von Humboldt
Foundation and is grateful for the kind hospitality of the Bethe
Center for Theoretical Physics at Bonn University. E.L.'s work was
partially supported by the STDF project 37272. S. Nasri thanks the
ICTP where part of this work was carried out.
M.Toharia would like to thank FRQNT for partial financial support
under grant numbers PRCC-191578 and PRC-290000.


\section{Appendix: Analytical Treatment of phase-broken $\mu-\tau$.}

Using Eqs.~(\ref{phasebrokenmutau0})-(\ref{M23}),  the constraints $|M_{e \mu}|=|M_{e \tau}|$ and $|M_{\mu \mu}|=|M_{\tau
  \tau}|$ can be recast as 
\bea
\label{M22M33:appendix}
M_{\mu \mu}&=& e^{2i (\theta-2\sigma)} M_{\tau \tau}\\
\label{M12M13:appendix}
M_{e \mu}&=&e^{2 i(\theta-\sigma)}M_{e \tau}.
\eea


The two complex equations (\ref{M22M33:appendix}) and
(\ref{M12M13:appendix}) of  ``phase breaking'' $\mu-\tau$ symmetry represent
two real constraint equations among the neutrino sector physical parameters,
and two real equations involving the phase breaking parameters $\theta$ and
$\sigma$ in terms of the neutrino physical parameters.
We cast them as
\begin{eqnarray}
(|M_{\mu \mu}|^2 -|M_{\tau \tau}|^2)& =&(|M_{e \tau}|^2 - |M_{e \mu}|^2)
  \ \ \ \ \ \  (``H_{22}=H_{33}\ equation'' )\
   \label{betad}\\
|M_{e \mu}|^2 &=& |M_{e \tau}|^2
 \ \ \ \ \ \ \ \ \ \ \ \ (``sum-rule\ equation'')    \label{sumrule}\\
e^{2 i\theta} &=&\left(\frac{M_{\tau\tau}M^2_{e\mu}}{M_{\mu\mu}M^2_{e\tau}}\right)
\ \ \ \ \ \ \ \ \ \ \ \ (``\theta\  \ equation'')            \label{theta}\\
e^{2 i \sigma}& =&\left(\frac{M_{\tau\tau}M_{e\mu}}{M_{\mu\mu}M_{e\tau}}\right)
\ \ \ \ \ \ \ \ \ \ \ \ (``\sigma \ \ equation'')              \label{phi}
\end{eqnarray}
where the matrix $H$ is defined as $H_{\nu}=M_{\nu}M_{\nu}^{\dagger}$.
The  ``$\sigma$ equation'' involves the unphysical phase $\sigma$ as
well as the unphysical diagonalization phases $\gamma_2$ and
$\gamma_3$ and thus does not constrain directly any physical
parameter.

However the other three equations have direct phenomenological effect
on the physical parameters. Note that $\theta$ is the only parameter coming from
the original neutrino mass matrix and  from Eq.~(\ref{theta}) we can extract  its value once we fix
all the observable parameters in the neutrino sector, i.e. $|V_{e3}|,|V_{\mu3}|,|V_{e2}|,
\delta, \eta, \xi, |m_1|,|m_2|$ and $|m_3|$.

Note that Eqs.~(\ref{betad})  and (\ref{sumrule}) represent
constraints among the physical parameters of the neutrino sector.
We choose to eliminate $|V_{\mu3}|$ using Eq.~(\ref{betad}) (which we
refer to as the ``$H_{22}=H_{33}$ equation'') and $\eta$
using equation (\ref{sumrule})  (which we call the ``$sum-rule$
equation'') and in particular we obtain the dependencies $|V_{\mu3}| \equiv f(|V_{e3}|,|V_{e2}|,
\delta,|m_1|,|m_2|,|m_3|)$ and $\eta \equiv g( |V_{e3}|,|V_{e2}|,
\delta, \xi, |m_1|,|m_2|,|m_3|)$. 
From these, the phase breaking parameter $\theta$ has a
dependency as $\theta \equiv h(|V_{e3}|,|V_{e2}|,
\delta, \xi, |m_1|,|m_2|,|m_3|)$ 
and so fixing $|V_{e3}|,|V_{e2}|,
\delta, |m_2|,|m_3|$ to their experimental values  gives us a simpler
parameter dependency $\theta \equiv h(\xi, |m_{lightest}|)$.
Before going into details on the main constraint equations from this
setp, we will show a revealing table with the counting of parameters
in the different implementations of $\mu-\tau$ symmetry discussed
here.
\begin{table}[h]
 \begin{center}
{\small
 \begin{tabular}{||c||c|c|c||c|c||}
\hline
\hline
 Setup& $M_\nu$ &  $D_\nu$ & $U_\nu$ & $V_{\mbox{\tiny{PMNS}}}$ & $P_L$ \\
  & &$\mbox{diag}(|m_1|,|m_2|,|m_3|)$\  &\ $U^\dagger_\nu M_\nu U^*_\nu = D_\nu$\ &\ $U_\nu= P_L V_{\mbox{\tiny{PMNS}}}$ &\\
\hline
General case                               & $12$ & $3$ & $9$ & $6$ & $3$\\ \hline
Exact $\mu-\tau$ permutation   & $8$    &  $3$ & $5$ & $3$ & $2$ \\ \hline
Exact $\mu-\tau$ reflection       &  6     &$3$ & $3$ & $2$ & $1$\\ \hline
Phase broken $\mu-\tau$           & $10$ & $3$ & $7$ & $4$ & $3$ \\
\hline\hline
 \end{tabular}
 }
 \end{center}
\vspace{-.4cm}
 \caption{ \label{tab1} Number of real free parameters for the
    neutrino mass related matrices, in the basis where the charged
    lepton mass matrix is diagonal and real, for different $\mu-\tau$
    symmetry implementations.} 
 \end{table}

\subsection{The ``$H_{22}=H_{33}$'' equation}

Let us begin with the general case which corresponds to $|M_{e \mu}|=|M_{e \tau}|$
and $|M_{\mu \mu}|=|M_{\tau \tau}|$. These two equations can be
combined to give the constraint
\begin{eqnarray}
  (|M_{\mu\mu}|^2 -|M_{\tau\tau}|^2)& =&(|M_{e\tau}|^2 -
  |M_{e\mu}|^2)\label{betd}
\eea
We then realize that at the level of the hermitian matrix $H_{\nu}=M_{\nu}M_{\nu}^{\dagger}$,  the two diagonal
elements $H_{22}$ and $H_{33}$ are given by
$H_{22}=|M_{e\mu}|^2+|M_{\mu\mu}|^2+|M_{\mu\tau}|^2$ and
$H_{33}=|M_{e\tau}|^2+|M_{\mu\tau}|^2+|M_{\tau\tau}|^2$. We see that
Eq.~(\ref{betd}) is equivalent to enforcing  $H_{22}=H_{33}$ at the level of the
hermitian matrix $H_\nu$.
Note that working with the elements of $H_\nu$ has the benefit of
removing the dependance on the Majorana phases $\eta$ and $\xi$. We
have in particular
\begin{eqnarray}
H_{22} &=& |m_1|^2+(|m_2|^2-|m_1|^2)|V_{\mu2}|^2+(|m_3|^2-|m_1|^2)|V_{\mu3}|^2  \\
H_{33} &=& |m_1|^2+(|m_2|^2-|m_1|^2)|V_{\tau 2}|^2+(|m_3|^2-|m_1|^2)|V_{\tau 3}|^2
\end{eqnarray}
Imposing the condition $H_{22}=H_{33}$ and using the definitions for
$V_{\mu2}$ and $V_{\tau 2}$ from Eqs.~(\ref{vmu2}) and (\ref{vtau2}) we obtain the expression
\begin{eqnarray}
X\left[(1-|V_{e3}|^2)^2-r[|V_{e1}|^2-|V_{e2}|^2|V_{e3}|^2]\right]=
rD\cos(\delta_{CP}) \label{h22h33NO:appendix}
\end{eqnarray}
for a normal neutrino mass spectrum ($\Delta_{atm}^2=|m_3|^2-|m_1|^2$) and
\begin{eqnarray}
X\left[(1-|V_{e3}|^2)^2+r[|V_{e1}|^2-|V_{e2}|^2|V_{e3}|^2]\right]= -rD\cos(\delta_{CP}) \label{h22h33IO:appendix}
\end{eqnarray}
for an inverted spectrum ($\Delta_{atm}^2=|m_1|^2-|m_3|^2$).

In both expressions $\Delta_{sol}^2=|m_2|^2-|m_1|^2$ and
$r=\frac{\Delta_{sol}^2}{\Delta_{atm}^2}$, and we have defined
\begin{eqnarray}
X &=& |V_{\tau3}|^2-|V_{\mu3}|^2 \\
D &=& 4|V_{e1}||V_{e2}||V_{e3}||V_{\mu 3}||V_{\tau3}|
\end{eqnarray}
in order to simplify the notation. From these equations one can obtain
an exact expression for the atmospheric neutrino mixing element given
by
\bea
|V_{\mu3}|^2_{NH} = \frac{1}{2}\left(1 - |V_{e3}|^2\right) \left(1 -
\frac{\varepsilon_{N}}{\sqrt{1+\varepsilon_{N}^2}} \right)
\eea
in the normal hierarchy regime, and where we have defined the (small) parameter $\varepsilon_{N}$ as
\bea
\varepsilon_{N} = \frac{2r|V_{e1}||V_{e2}||V_{e3}|
  \cos(\delta_{CP})}{  (1-|V_{e3}|^2)^2-r[|V_{e1}|^2-|V_{e2}|^2|V_{e3}|^2]}.
\eea
For the inverted hierarchy regime, we obtain
\bea
|V_{\mu3}|^2_{IH} = \frac{1}{2}\left(1 - |V_{e3}|^2\right) \left(1 +
\frac{\varepsilon_{I}}{\sqrt{1+\varepsilon_{I}^2}} \right)
\eea
and again we have defined the (small) parameter $\varepsilon_{I}$ as
\bea
\varepsilon_{I} = \frac{2r|V_{e1}||V_{e2}||V_{e3}|
  \cos(\delta_{CP})}{  (1-|V_{e3}|^2)^2 + r[|V_{e1}|^2-|V_{e2}|^2|V_{e3}|^2]}.
\eea

By performing an expansion in powers of $\varepsilon_{N}$ and
$\varepsilon_{I}$ and keeping the lowest terms one trivially reproduces the
expressions given in the main text in Eqs.~(\ref{vmu3NO}) and (\ref{vmu3IO}).

\subsection{The ``Sum-Rule'' Equation}

To extract useful information on the ${\cal CP}$ violating phase $\delta$, it is more convenient to express
the second constraint, $|M_{e\mu}|^2=|M_{e\tau}|^2$ as
\begin{eqnarray}
X\left[F_1+|V_{e3}|^2F_2\right]+D\left[\tilde{K}_1\cos(\delta)+\tilde{K}_2\sin(\delta)\right]=0
\end{eqnarray}
where we have defined
\bea
F_1&=&|V_{e1}|^2|V_{e2}|^2\left[|m_1|^2+|m_2|^2-2|m_1||m_2|\cos(2\eta)\right]\\
F_2&=&2|m_1||m_3||V_{e1}|^2(1-|V_{e3}|^2)\cos(2\xi-2\delta)
+2|m_2||m_3||V_{e2}|^2(1-|V_{e3}|^2)\cos(2\eta-2\xi+2\delta) \non\\
&&-2|m_1||m_2||V_{e1}|^2|V_{e2}|^2\cos(2\eta)-|m_1|^2|V_{e1}|^4-|m_2|^2|V_{e2}|^4-|m_3|^2(1-|V_{e3}|^2)^2\\
\tilde{K}_1 &=&|m_2|^2|V_{e2}|^2-|m_1|^2|V_{e1}|^2-2|m_1||m_2||V_{e2}|^2\cos(2\eta)
\non \\
&&+\left[|m_1||m_2|\cos(2\eta)+|m_1||m_3|\cos(2\xi)-|m_2||m_3|\cos(2\eta-2\xi)\right](1-|V_{e3}|^2)\\
\tilde{K}_2 &=& \left[|m_1||m_2|\sin(2\eta)+|m_1||m_3|\sin(2\xi)+|m_2||m_3|\sin(2\eta-2\xi)\right](1-|V_{e3}|^2).
\eea
Injecting the value of the parameter $X$ obtained from the
``$H_{22}=H_{33}$ equation'' (Eqs.~(\ref{h22h33NO:appendix}) and
(\ref{h22h33IO:appendix}) )  we can eliminate the dependance on $|V_{\mu3}|$ and
obtain, for the normal mass hierarchy,
\begin{eqnarray}
  \label{sumrulenormal}
\cot(\delta)_{NH} = -\frac{(1-|V_{e3}|^2)^2(1-rZ)
  \tilde{K}_2}{(1-|V_{e3}|^2)^2(1-rZ) \tilde{K}_1+r\left[F_1+|V_{e3}|^2F_2\right]},
\end{eqnarray}
and for the inverted mass hierarchy,
\begin{eqnarray}
  \label{sumruleinverted}
\cot(\delta)_{IH} = -\frac{(1-|V_{e3}|^2)^2(1+rZ)\tilde{K}_2}{(1-|V_{e3}|^2)^2(1+rZ)\tilde{K}_1-r\left[F_1+|V_{e3}|^2F_2\right]}
\end{eqnarray}
where,
\begin{eqnarray}
Z &=& \frac{|V_{e1}|^2-|V_{e2}|^2|V_{e3}|^2 }{(1-|V_{e3}|^2)^2}.
\end{eqnarray}
The above equations depend only on the physical parameters
$|m_1|$, $|m_2|$, $|m_3|$, $\delta_{CP}$, $\xi$, $\eta$,  $|V_{e2}|$
and $|V_{e3}|$.
Note also that the expression $F_2$ contains some further dependence on $\delta_{CP}$. However,
if we drop terms proportional to $r$ and $|V_{e3}|^2$, then we obtain a
simple relation between $\cot(\delta)$ and the neutrino masses, the
$|V_{ij}|^2$ and the two Majorana ${\cal CP}$ violating phases $\eta$
and $\xi$
\begin{eqnarray}
\cot(\delta) = -\frac{K_2}{K_1} \ \ +\ {\cal O}(r,|V_{e3}|^2)
\end{eqnarray}
valid for both normal and inverted mass orderings and with $K_{i}$
(without tilde)  defined as
\bea
K_1 &=&|m_2|^2|V_{e2}|^2-|m_1|^2  (1-|V_{e2}|^2) + (1 -2  |V_{e2}|^2)
|m_1||m_2| \cos(2\eta)\non\\
 &&\ +|m_1||m_3|\cos(2\xi)-|m_2||m_3|\cos(2\eta-2\xi)\\
K_2 &=&
|m_1||m_2|\sin(2\eta)+|m_1||m_3|\sin(2\xi)+|m_2||m_3|\sin(2\eta-2\xi) .
\eea

\subsection{The ``$\theta$'' equation}

From the {\it phase-broken $\mu-\tau$} symmetry constraints
we can obtain an expression of $\tan(\theta)$ as
\bea
  \tan(\theta)&=&\frac{Im\left[M_{\mu \mu}^*M_{\tau \tau}(M_{e \mu}M_{e \tau}^*)^2\right]}{|M_{\mu \mu}|^2|M_{e \mu}|^4+Re\left[M_{\mu \mu}^*M_{\tau \tau}(M_{e \mu}M_{e \tau}^*)^2\right]}
\end{eqnarray}
where the $M_{ij}$ are defined in Eqs.~(\ref{M22})-(\ref{M13}).
The exact expression of $\tan(\theta)$ in terms of masses and mixing
angles is long and not very revealing. However, the approximate
relations keeping only the lowest terms in $|V_{e3}|$ and $r$,  and in the special limits of $|m_1|=0$ and
$|m_3|=0$ are simple and match the numerical results obtained using the
exact expression.
\bit

\item In the case of $|m_1|=0$ and expanding in $|V_{e3}|$ and $r$ we
  have,  to lowest order,
\begin{eqnarray}
\tan(\theta)\simeq \frac{2\sqrt{r}|V_{e1}||V_{e2}||V_{e3}|}{\left[r|V_{e1}|^2|V_{e2}|^2-|V_{e3}|^2\right]}\sin(2\eta-2\xi+\delta)
\end{eqnarray}
However we still need to implement in the above expression the
condition coming from the sum rule constraint from Eq~(\ref{sumrulenormal}), which in this limit can be written as,
\begin{eqnarray}
\sin(2\eta-2\xi+\delta)&=& \pm\left[1-\frac{r}{2}|V_{e2}|^4\cos^2(\delta)\right]
\end{eqnarray}
Injecting the above value in the expression of $\tan(\theta)$, we finally obtain 
\begin{eqnarray}
\tan(\theta)\simeq \pm \frac{2\sqrt{r}|V_{e1}||V_{e2}||V_{e3}|}{\left[r|V_{e1}|^2|V_{e2}|^2-|V_{e3}|^2\right]}
\end{eqnarray}

\item The  case $|m_3|=0$ is more subtle. Again, for small $|V_{e3}|$, we obtain,
\begin{eqnarray}
\tan(\theta) = \frac{S}{2R}+\frac{Q}{P}\  +\  {\cal O}(|V_{e3}|^3)
\end{eqnarray}
where,
\begin{eqnarray}
  S &=& -2 |V_{e1}||V_{e2}||V_{e3}| \sin(\delta) \left( 4
  (1-2|V_{e2}|^2) \sin^2(\eta) + r \right)\\
Q &=& S + 4 r  |V_{e1}||V_{e2}||V_{e3}| \sin(\delta) \\
P &=& 8 |V_{e2}|^2(1-|V_{e2}|^2)\sin^2(\eta)-2\left[1+ 4 |V_{e2}|^2\sin^2(\eta)\right]|V_{e3}|^2\\
R&=&1- 4 |V_{e2}|^2(1-|V_{e2}|^2)\sin^2(\eta)-r(1-|V_{e2}|^2)-4\left[\cos^2(\eta)-\cos^2(\delta)\cos(2\eta)\right]|V_{e3}|^2 \\
&& -4\sin^2(\eta)|V_{e2}|^2\left[1-2|V_{e2}|^2+4(1-|V_{e2}|^2)\sin^2(\delta)\right]|V_{e3}|^2+2(1-2|V_{e2}|^2)\sin(2\delta)
\sin(2\eta)|V_{e3}|^2. \nonumber
\end{eqnarray}
Two different limits for $\eta$ stand out from the previous expression. 
\bit
\item For $\sin(\eta) \ll |V_{e3}|$, the expression
simplifies to
\begin{eqnarray}
\tan(\theta)&\simeq& -\frac{r|V_{e1}||V_{e2}|}{|V_{e3}|}\sin(\delta)
\end{eqnarray}
\item For $\sin(\eta) \gg |V_{e3}|$, the expression becomes
\begin{eqnarray}
\tan(\theta)&\simeq&
-\frac{\left[1-4|V_{e1}|^2|V_{e2}|^2\cos^2(\delta)\right]|V_{e3}|\sin(\delta)}{|V_{e1}||V_{e2}|\left[1-2|
   V_{e2}|^2\right]}
\end{eqnarray}
 \item For $\sin(\eta) \simeq  {\cal O} (|V_{e3}|)$ the analytical
   expression is cumbersome and not specially revealing.  
\eit

  \eit

\end{document}